\documentclass[preprint,aps,pre,eqsecnum,showpacs,showkeys,a4paper,longbibliography]{revtex4-1}  

\usepackage{hyperref} 
\usepackage{amsmath}
\usepackage{amssymb}

\renewcommand{\d}{{\,\rm d}}
\newcommand{\e}{{\rm e}}
\newcommand{\p}{{\partial}}
\newcommand{\VEC}[1]{{\mathbf{#1}}}

\newcommand{\intlim}{\int\limits}
\newcommand{\meant}[1]{\left<#1\right>}
\newcommand{\meanx}[1]{\overline{#1}}

\begin{document}

\title{A classical explanation of quantization}% Force line breaks with \\

\author{Gerhard \surname{Gr\"ossing}}
\email[E-mail: ]{ains@chello.at}
\homepage{http://www.nonlinearstudies.at/}
\author{Johannes \surname{Mesa Pascasio}}
\email[E-mail: ]{ains@chello.at}
\homepage{http://www.nonlinearstudies.at/}
\author{Herbert \surname{Schwabl}}
\email[E-mail: ]{ains@chello.at}
\homepage{http://www.nonlinearstudies.at/}
\affiliation{%
Austrian Institute for Nonlinear Studies\\
Akademiehof, 
Friedrichstr.~10, 1010 Vienna, Austria
\vspace*{8.0cm}
}

\begin{abstract}
In the context of our recently developed emergent quantum mechanics, and, in particular, based on an assumed sub-quantum thermodynamics, the necessity of energy quantization as originally postulated by Max Planck is explained by means of purely classical physics. Moreover, under the same premises, also the energy spectrum of the quantum mechanical harmonic oscillator is derived. Essentially, Planck's constant $h$ is shown to be indicative of a particle's ``zitterbewegung'' and thus of a fundamental angular momentum. The latter is identified with quantum mechanical spin, a residue of which is thus present even in the non-relativistic Schr\"odinger theory.
\end{abstract}
                 
\maketitle 

\section{Introduction}
\label{sec:intro} 

In references \cite{Groessing.2008vacuum,Groessing.2009origin}, the Schr\"odinger equation was derived in the context of modelling quantum
systems via nonequilibrium thermodynamics, i.e., by the requirement that the
dissipation function, or the time-averaged work over the system of interest, vanishes
identically. The ``system of interest'' is a ``particle'' 
embedded in a thermal environment of non-zero average temperature, i.e.,
of the vacuum's zero-point energy (ZPE).
Ours is thus an approach in the tradition of stochastic mechanics (see, e.g., 
Nelson \cite{Nelson.1966}, 
Fritsche and Haugk \cite{Fritsche.2003}, 
Guerra \cite{Guerra.1984}), or 
stochastic electrodynamics, respectively (see, e.g., 
de~la~Pe\~na and Cetto \cite{Pena.1996}, 
Boyer \cite{Boyer.1980}, 
Haisch et~al. \cite{Haisch.1994}).
In more recent papers \cite{Groessing.2010emergence,Groessing.2010entropy,Groessing.2010free}, we have modelled the ``particle'' more
concretely by using an analogy to the ``bouncer'' gleaned from the
beautiful experiments by Couder's group \cite{Couder.2005,Couder.2006,Protiere.2006,Eddi.2009,Fort.2010}. 
This analogy is here expanded to a ``particle'' moving in three dimensions, which is denoted as ``walker''.
One assumes that the thermal ZPE
environment is oscillating itself, with the kinetic energy of these latter oscillations
providing the energy necessary for the ``particle'' to maintain a constant energy, i.e., to
remain in a nonequilibrium steady-state.
Referring to the respective (zero-point) oscillations of the vacuum, one 
assumes the particle oscillator to be embedded in an environment comprising a corresponding energy bath with both periodic and stochastic contributions.

In the present paper we discuss in detail this two-fold perspective of an individual ``particle'', i.e., comprising 
a first one where it is imagined as an oscillating ``bouncer'', and
a second one where it is considered as a ``walker'' which performs a stochastic movement in three dimensions. 
After individual inspection, these two tools will be compared, or coupled, respectively.
After all, as the experiments of Couder's group show so impressively, a ``particle'' may both oscillate in time in a regular fashion 
(i.e., with a characteristic frequency) \textit{and} propagate irregularly in space (i.e., via Brownian-type motion).
With respect to our analogous sub-quantum model, then, this means that the ``zitterbewegung'' is simultaneously characterized by regular periodic and stochastic motions, both of which must, however, be comparable on the level of the work-energy expended during a certain amout of time.
In Chapters~\ref{sec:aceq.sub.3} and \ref{sec:sub.2} we shall calculate the respective work-energies for each aspect separately, 
whereas in Chapter~\ref{sec:walking} they will be compared during one and the same time-span.
This will lead to Planck's quantization relation.

\section{A classical oscillator driven by its environment's energy bath: the ``bouncer''}
\label{sec:aceq.sub.3}

Let us start with
the following Newtonian equation for a classical oscillator with one degree of freedom (DOF)
\begin{equation}
    m\ddot{x} = -m\omega_0^2x - 2\gamma m\dot{x} + F_0\cos\omega t\;.  \label{eq:2.1}
\end{equation}
Eq.~(\ref{eq:2.1}) describes a forced oscillation of a mass $m$ swinging around a center point along $x(t)$ with amplitude $A$ and damping factor, or friction, $\gamma$. If $m$ could swing freely, its resonant angular frequency would be $\omega_0$. Due to the 
damping of the swinging particle there is a need for a locally independent driving force $F(t) = F_0\cos\omega t$.

We are only interested in the stationary solution of Eq.~(\ref{eq:2.1}), i.e., for $t\gg\gamma^{-1}$, where $\gamma^{-1}$ plays the role of a relaxation time, using the ansatz
\begin{equation}
	x(t) = A\cos(\omega t + \varphi)\;.  \label{eq:2.2}
\end{equation}
One finds for the phase shift between the forced oscillation and the forcing oscillation that
\begin{equation}
	\tan\varphi = -\frac{2\gamma\omega}{\omega_0^2 - \omega^2}\;,  \label{eq:2.3}
\end{equation}
and for the amplitude of the forced oscillation
\begin{equation}
	A(\omega) = \frac{F_0/m}{\sqrt{(\omega_0^2 - \omega^2)^2 + (2\gamma\omega)^2}}\;. 	\label{eq:2.4}
\end{equation}

To analyse the energetic balance, one multiplies Eq.~(\ref{eq:2.1}) with $\dot{x}$ and obtains
\begin{align}
  m\ddot{x}\dot{x} + m\omega_0^2x \dot{x} = -2\gamma m\dot{x}^2 + F_0\cos(\omega t)\dot{x}\;,  \label{eq:2.5}
\end{align}
and thus,
\begin{align}
  \frac{\rm d}{{\rm d}t}\left(\frac{1}{2}m \dot{x}^2 + \frac{1}{2} m\omega_0^2x^2\right) 
     = -2\gamma m\dot{x}^2 + F_0\cos(\omega t)\dot{x} = 0\;.  \label{eq:2.6}
\end{align}
The Hamiltonian of the system is the term within the brackets, %of the left-hand side of Eq.~(\ref{eq:2.6}),
\begin{equation}
  \mathcal{H} = \frac{1}{2}m \dot{x}^2 + \frac{1}{2} m\omega_0^2 x^2 = \text{const.},  \label{eq:2.14}
\end{equation}
thus providing the vanishing of Eq.~\eqref{eq:2.6}.

Due to the friction the oscillator looses its energy to the bath, viz., the power term represented by $-2\gamma m\dot{x}^2$, whereas $F_0\cos(\omega t)\dot{x}$ represents the power which is regained from the energy bath via the force $F(t)$. 
As the sum of the two terms of \eqref{eq:2.6} is zero, one can write down the net work-energy that is taken up by the bouncer during each period $\tau$ as
\begin{align}
  W_{\rm bouncer} &= \intlim_\tau F_0 \cos(\omega t)\dot{x} \d t = \intlim_\tau 2\gamma m\dot{x}^2 \d t \nonumber\\  
	  &= 2\gamma m\omega^2 A^2\intlim_\tau \sin^2(\omega t + \varphi)\d t  \nonumber\\
	  &= \gamma m\omega^2A^2\tau\;. \label{eq:2.7}
\end{align}

To derive the stationary frequency $\omega$, we use the right-hand side of Eq.~(\ref{eq:2.6}) together with Eq.~(\ref{eq:2.2}) to first obtain
\begin{align}
  2\gamma m\dot{x} 
  = -2\gamma m A\omega\sin(\omega t + \varphi) 
  = F_0\cos\omega t\;.  \label{eq:2.8}
\end{align}
As all factors, except for the sinusoidal ones, are time independent, we have the necessary condition for the phase given by
\begin{equation}
  -\sin(\omega t + \varphi) = \cos\omega t  \quad\Rightarrow\quad \varphi = -\frac{\pi}{2} + 2n\pi  \label{eq:2.9}
\end{equation}
for all $n\in \mathbb{Z}$. Substituting this into Eq.~(\ref{eq:2.3}), we obtain
\begin{equation}
  \tan\left(-\frac{\pi}{2} + 2n\pi\right) = \pm\infty = -\frac{2\gamma\omega}{\omega_0^2 - \omega^2}\;,  \label{eq:2.10}
\end{equation}
and thus
\begin{equation}
  \omega = \omega_0\;.  \label{eq:2.11}
\end{equation}
Therefore, {\it the system turns out to be stationary at the resonance frequency $\omega_0$ of the free undamped oscillator}. With the notations
\begin{equation}
	\tau = \frac{2\pi}{\omega_0}\;,\quad r := A(\omega_0) = \frac{F_0}{2\gamma m\omega_0}\;,  \label{eq:2.12}
\end{equation}
we obtain
\begin{equation}
  W_{\rm bouncer} = W_{\rm bouncer}(\omega_0) = \gamma m\omega_0^2r^2\tau = 2\pi \gamma m\omega_0r^2\;. \label{eq:2.13}
\end{equation}

If one introduces the angle $\theta(t) := \omega_0t$ and substitutes Eq.~(\ref{eq:2.2}) into Eq.~(\ref{eq:2.14}), this yields, as is well known, the two equations
\begin{align}
  \ddot{r} - r\dot{\theta}^2 + \omega_0^2r &= 0\;,  \label{eq:2.15}
\end{align}
and
\begin{align}
  r\ddot{\theta} + 2\dot{r}\dot{\theta} &= 0\;. \label{eq:2.16}
\end{align}
From Eq.~(\ref{eq:2.16}), an invariant quantity is obtained: it is the angular momentum,
\begin{equation}
  L(t) = mr^2\dot{\theta}(t)\;.  \label{eq:2.17} 
\end{equation}
With $\theta(t) = \omega_0t$, and thus $\dot{\theta}=\omega_0$, the quantity of Eq.~(\ref{eq:2.17}) becomes
a time-invariant expression, which we denote as
\begin{equation}
  \hbar := mr^2\omega_0\;.  \label{eq:2.18}
\end{equation}
Note that $L(t)$ is an invariant even more generally, i.e., for $\theta(t):=\int\omega(t)\d t$. Still, for all cases where the time average $\meant{\theta(t)} =\omega_0 t$, one can again write down $\hbar$ in the form Eq.~\eqref{eq:2.18}.
Thus, we rewrite our result (\ref{eq:2.13}) as
\begin{equation}
  W_{\rm bouncer} = 2\pi\gamma\hbar\;.  \label{eq:2.19}
\end{equation}

For the general case of $N$ dimensions, we make use of Eq.~\eqref{eq:2.2} independently in any of the $N$ directions,
\begin{equation} \label{eq:2.20}
  \begin{array}{rcl}
    x_1(t) &=& A_{x_1}\cos(\omega_0 t + \phi_{x_1}) \;, \\
     &\vdots&    \\
    x_N(t) &=& A_{x_N}\cos(\omega_0 t + \phi_{x_N}) \;, \\
  \end{array}
\end{equation}
with the same frequency $\omega_0$ in any direction as was obtained in \eqref{eq:2.11}. 
Moreover, replacing $r$ in Eq.~\eqref{eq:2.12} by its $N$-dimensional version $\VEC{r}$, with the corresponding coupled Eqs.~\eqref{eq:2.15} and \eqref{eq:2.16} for $\VEC{r}$, provides our time invariant expression as
\begin{equation}  \label{eq:2.20a}
  \hbar = m\omega_0 \VEC{r}\cdot\VEC{r} \;.
\end{equation}
As we can treat each direction independently, we obtain $N$ components of the work-energy during each period $\tau$,
\begin{align}  
  W_{\rm bouncer} &= \intlim_\tau 2\gamma m(\dot{x_1}^2 + \cdots + \dot{x_N}^2) \d t 
	  = \gamma m\omega^2A^2\tau\;. \label{eq:2.21}
\end{align}
Thus, it holds also for any number $N$ of dimensions that
\begin{equation}
  W_{\rm bouncer} = 2\pi\gamma\hbar\;.  \label{eq:2.22}
\end{equation}

\section{Brownian motion of a particle: the ``walker''}
\label{sec:sub.2}

In a second step, we introduce a ``particle'' driven via a stochastic force,
e.g., due to not just one regular, but to different fluctuating wave-like configurations in the environment. Therefore, our ``particle's'' motion will generally assume a Brownian-type character. 
The Brownian motion of a thus characterized particle, which we propose to call a ``walker'', is then described (in any one dimension) by a Langevin stochastic differential equation with 
velocity $u=\dot{x}$, force $f(t)$, and friction coefficient $\zeta$,
\begin{equation}
  m\dot{u} = -m\zeta u + f(t)\;,  \label{eq:3.1}
\end{equation}
The time-dependent force $f(t)$ is stochastic, i.e., one has as usual for the time-averages
\begin{equation}
  \meant{f(t)} = 0\;,\quad \meant{f(t)f(t')} = \phi(t-t')\;,  \label{eq:3.2}
\end{equation}
where $\phi(t)$ differs noticeably from zero only for $t < \zeta^{-1}$. The correlation time $\zeta^{-1}$ denotes the time during which the fluctuations of the stochastic force remain correlated.

In analogy to the standard textbook solution for Eq.~(\ref{eq:3.1}) in terms of the mean square displacements $\meanx{x^2}$ as given
from Ornstein-Uhlenbeck theory \cite{Coffey.2004}, one can write
\begin{equation}
  \meanx{x^2} = \frac{4E_{\rm zp}}{\zeta^2m}\left(\zeta\lvert t\rvert-1+\e^{-\zeta\lvert t\rvert}\right)\;,  \label{eq:3.3}
\end{equation}
where we introduce the average kinetic Energy $E_{\rm zp}$ of the zero-point field.
One can define the $E_{\rm zp}$ per DOF as
\begin{equation}
  E_{\rm zp} = \frac{kT_0}{2} \;,  \label{eq:3.3a}
\end{equation}
which is the sub-quantum analogon to the thermodynamical expression $k_{\rm B}T/2$, where $k_{\rm B}$ is Boltzmann's constant, and $T$ the classical
temperature, whereas $T_0$ in our scenario denotes the vacuum temperature. However, as we today neither know $T_0$ nor the constant $k$ 
(unless it should turn out as identical to $k_{\rm B}$), we shall mostly stick to formally using $E_{\rm zp}$. In other words, we shall use the
specification ``$kT_0$'' only occasionally, i.e., in order to point out the close analogy to the usual thermodynamical formalism, and as a reminder
that $E_{\rm zp}$ is the ``kinetic temperature'' of the vacuum's heat reservoir.

We stress that even if we use the same character $x$ as for the oscillating particle, now the meaning is 
different: $x(t)$ in the previous Chapter signified a deterministic harmonic displacement of mass point 
$m$ in the case of an oscillating particle (``bouncer''), whereas $x(t)$ now means a stochastic 
random walk variable for the particle that carries out a Brownian motion (``walker'').
Note that, on the one hand, for $t\ll\zeta^{-1}$, and by expanding the exponential up to second order, Eq.~(\ref{eq:3.3}) provides that
\begin{equation}
  \meanx{x^2} = \frac{2E_{\rm zp}}{m}t^2 = \frac{mu_0^2}{m}t^2 = u_0^2t^2\;,  \label{eq:3.4}
\end{equation}
with $u_0$ being the initial velocity fluctuation \cite{Groessing.2010emergence}.

On the other hand, for $t\gg\zeta^{-1}$, one obtains the familiar relation for Brownian motion, i.e., in one dimension,
\begin{equation}
  \meanx{x^2} \simeq 2Dt\;,  \label{eq:3.5}
\end{equation}
with the ``diffusion constant'' $D$ given by
\begin{equation}
  D = \frac{2E_{\rm zp}}{\zeta m}\;.  \label{eq:3.6}
\end{equation}
In $N$ dimensions, Eq.~\eqref{eq:3.5} would read $\meanx{x^2} \simeq 2NDt$, with the same $D$ as in Eq.~\eqref{eq:3.6}.

To obtain a better understanding of Equations (\ref{eq:3.5}) and (\ref{eq:3.6}), we want to detail here, in one dimension for simplicity, 
how they come about.
One usually introduces a coefficient $\lambda$ that measures the strength of the mean square deviation of the stochastic force, such that
\begin{equation}
  \phi(t) = \lambda\delta(t)\;.  \label{eq:3.7}
\end{equation}
Since friction increases in proportion to the frequency of the stochastic collisions, there must exist a connection between 
$\lambda$ and $\zeta$. One solves the Langevin equation (\ref{eq:3.1}) in order to find this connection. 
Solutions of this equation are well known from the Ornstein-Uhlenbeck theory of Brownian motion \cite{Coffey.2004}.

Since the dependence of $f(t)$ is known only statistically, one does not consider the average value of $u(t)$, but instead that of its square,
\begin{equation}
  \begin{aligned}
    \meanx{u^2(t)} 
     &= \e^{-2\zeta t}\intlim_0^t \d\tau\,\intlim_0^t\d\tau'\e^{\zeta(\tau+\tau')} \phi(\tau-\tau')\frac{1}{m} + u_0^2\e^{-2\zeta t}  \\
     &= \frac{\lambda}{2\zeta m^2}\left(1-\e^{-2\zeta t}\right) + u_0^2\e^{-2\zeta t}\quad 
        \stackrel{t\gg\zeta^{-1}}{\longrightarrow} \quad \frac{\lambda}{2\zeta m^2}\;,  
  \end{aligned} \label{eq:3.8}
\end{equation}
with $u_0$  being the initial value of the velocity. For $t\gg\zeta^{-1}$, the term with $u_0$ becomes negligible, i.e.,   
$\zeta^{-1}$ then plays the role of a relaxation time. We require that our particle attains thermal equilibrium \cite{Groessing.2008vacuum,Groessing.2009origin}
after long times so that due to the \textit{equipartition theorem on the sub-quantum level} the average value of the kinetic energy becomes 
\begin{equation}
  \frac{1}{2}m \, \meanx{u^2(t)} = E_{\rm zp} = \frac{1}{2}kT_0\;.  \label{eq:3.9}
\end{equation}
Combining Eqs.~(\ref{eq:3.8}) and (\ref{eq:3.9}), one obtains the Einstein-type relation 
\begin{equation}
  \lambda = 4\zeta mE_{\rm zp} \;.  \label{eq:3.10}
\end{equation} 
Similarly, one obtains the mean square displacement of $x(t)$ for $t\gg\zeta^{-1}$. 
Therefore, one integrates twice to obtain the confirmation of our result (\ref{eq:3.5}), i.e.,
\begin{equation}
  \meanx{x^2(t)} = \intlim_0^t\d\tau \intlim_0^t\d\tau' \frac{\lambda}{2\zeta m^2}\e^{-\zeta\lvert \tau-\tau'\rvert} 
    \simeq \frac{\lambda}{\zeta^2 m^2}t = 2Dt\;, \label{eq:3.11}
\end{equation}
with the diffusion constant turning out as identical to Eq.~(\ref{eq:3.6}), i.e.,
\begin{equation}
  D = \frac{\lambda}{2\zeta^2m^2} = \frac{2E_{\rm zp}}{\zeta m}\;.  \label{eq:3.12}
\end{equation}

Now we remind ourselves that we have to do with a steady-state system. Just as with the friction $\zeta$ 
there exists a flow of (kinetic) energy into the environment, there must also exist 
a work-energy flow back into our system of interest. For its calculation, we multiply Eq.~(\ref{eq:3.1}) 
by $u=\dot{x}$ and obtain an energy-balance equation. With a natural number 
$n>0$ chosen so that $n\tau$ is large enough to make all fluctuating contributions negligible,
it yields for the duration of time $n\tau$ the net work-energy of the walker
\begin{equation}
  W_{\rm walker} = \intlim_{n\tau} m\zeta \, \meanx{\dot{x}^2} \d t = m\zeta \intlim_{n\tau} \meanx{u^2(t)} \d t\;.  \label{eq:3.13}
\end{equation}
Inserting (\ref{eq:3.9}), we obtain
\begin{equation}
  W_{\rm walker} = n\tau m\zeta \, \meanx{u^2(t)}
     = 2n\tau \zeta E_{\rm zp} \;.  \label{eq:3.14}
\end{equation}
In order to make the result comparable with Eq.~(\ref{eq:2.19}), we choose
$\tau=2\pi/\omega_0$ to be identical with the period of Eq.~\eqref{eq:2.12}.
The work-energy for the particle undergoing Brownian motion can thus be written as 
\begin{equation}
  W_{\rm walker} = n\frac{4\pi}{\omega_0}\zeta E_{\rm zp}\;. \label{eq:3.15}
\end{equation}

Turning now to the $N$-dimensional case,
the average squared velocity of a particle is
\begin{equation}  \label{eq:3.16}
  \meant{u^2} = \meant{u_{x_1}^2} + \cdots + \meant{u_{x_N}^2} \;,
\end{equation}
with equal probability for each direction, 
\begin{equation}  \label{eq:3.17}
  \meant{u_{x_1}^2} = \cdots = \meant{u_{x_N}^2} = \frac{1}{N}\meant{u^2}\;.
\end{equation}
Accordingly, the average kinetic energy of a moving particle with $N$ DOF becomes
\begin{equation}  \label{eq:3.18}
  E_{\rm zp}^{(N)} = \frac{1}{2}m\meant{u^2} = NE_{\rm zp}
\end{equation}
and thus 
\begin{equation}  \label{eq:3.19}
  \meant{u^2(t)} = 2N \, \frac{E_{\rm zp}}{m}\;.  
\end{equation}
Again, we note that Eq.~\eqref{eq:3.19} describes an energy equipartition which, however, here relates to the sub-quantum level, i.e., to the vacuum temperature $T_0$.
It should thus not be confused with the equipartition theorem as discussed, e.g., with respect to blackbody radiation and the Planck spectrum.

With the analogical explanation as for the one-dimensional case, we find for the work-energy of the walker in $N$-dimensional space
\begin{align}
    W_{\rm walker} 
      &= m\zeta \intlim_{n\tau} \left[\meant{u_{x_1}^2(t)} + \cdots + \meant{u_{x_N}^2(t)}\right] \d t 
       = m\zeta \intlim_{n\tau} \meant{u^2(t)} \d t\;.  \label{eq:3.20}
\end{align}
Inserting \eqref{eq:3.19}, we obtain
\begin{equation}
  W_{\rm walker} = n\tau m\zeta \meant{u^2(t)}
     = 2n\tau \zeta NE_{\rm zp}  \;,  \label{eq:3.21}
\end{equation}
which is $N$ times the value of the one-dimensional case in Eq.~\eqref{eq:3.15}. Therefore, the work-energy for the particle undergoing Brownian motion can be written as 
\begin{equation}
  W_{\rm walker} = n\frac{N 4\pi}{\omega_0}\zeta E_{\rm zp}\;, \label{eq:3.22}
\end{equation}
for the general case of $N$ DOF.

\section{\texorpdfstring{The ``walking bouncer'': Derivation of $E=\hbar\omega$}{Walking bouncer}}
\label{sec:walking}

We have so far analyzed two perspectives for our model of a single-particle quantum system:
\begin{enumerate}
	\item A harmonic oscillator is driven by the environment via a periodic force $F_0\cos\omega_0t$. In the center of mass frame, the system is characterized by a single DOF.
    However, in the $N$-dimensional reference frame of the laboratory,
    the oscillation is not fixed \textit{a priori}. Rather, with $\hbar$ as
    angular momentum, there will be a free rotation in all $N$ dimensions, and possible exchanges of energy will be equally distributed in a stochastic manner.
  \item Concerning the latter, the flow of energy is on average distributed evenly via the friction $\gamma$ in all
    $N$ dimensions of the laboratory frame. It can thus also be considered as the stochastic source of the particle moving in $N$ dimensions, each
    described by the Langevin equation \eqref{eq:3.1}.
\end{enumerate}
Accordingly, the walker gains its energy from the heat bath via the oscillations of the bouncer-bath system in $N$ dimensions:
The bouncer pumps energy to and from the heat bath via the ``friction'' $\gamma$. 
There is a continuous flow from the bath to the oscillator, and \textit{vice versa}. 
Therefore, we recognize ``friction'' 
in both cases, as represented by $\gamma$ and $\zeta$, respectively, to generally describe the coupling between the oscillator (or particle in motion) on the one hand, and the bath on the other hand.
Moreover, and most importantly, during that flow, 
for long enough times $n\tau$, this coupling of the bouncer can be assumed to be exactly identical with the coupling of the walker. For this reason we directly compare the results of Eqs.~(\ref{eq:2.22}) and (\ref{eq:3.22}),  
\begin{align}
  nW_{\rm bouncer} = W_{\rm walker}\;,  \label{eq:4.1}
\end{align}
providing
\begin{align}
  n2\pi\gamma\hbar = n\frac{N4\pi}{\omega_0}\zeta E_{\rm zp} \;,  \label{eq:4.2}
\end{align}
with $n\gg 1$ since we have to take the mean over a large number of stochastic motions.

Now, one generally has that the total energy of a sinusoidal oscillator exactly equals twice its average kinetic energy. Moreover, despite having a nonequilibrium framework of our system, the fact that we deal with a steady-state means that our oscillator is in local thermal equilibrium with its environment. 
As the average kinetic energy of the latter may be written as $E_{\rm zp}=kT_0/2$ for each degree of freedom, 
one has for the corresponding total energy that $E_{\rm tot}=2E_{\rm zp}^{(N)}=2NE_{\rm zp}=NkT_0$. 
Now, one can express that energy via Eq.~(\ref{eq:4.2}) in terms of the oscillator's frequency $\omega_0$, and one obtains for $N$ DOF
\begin{align}
  E_{\rm tot} = 2E_{\rm zp}^{(N)} = \frac{\gamma}{\zeta}\hbar\omega_0\;.  \label{eq:4.3}
\end{align}

To describe the steady-state's couplings in both systems, it is appropriate to assume
the same friction coefficient for both the bouncer and the walker, i.e., $\gamma=\zeta$. We obtain the energy balance between oscillator and its thermal environment
for $N$ DOF as
\begin{align}
  2E_{\rm zp}^{(N)} = \hbar\omega_0\;,  \label{eq:4.4}
\end{align}
i.e., it holds in particular in any single dimension that
\begin{align}
  2E_{\rm zp} = kT_0 = \hbar\omega_0  \;.  \label{eq:4.4a}
\end{align}
The total energy of our model for a quantum ``particle'', i.e., a driven steady-state oscillator system, is thus {\it derived} as
\begin{align}
  E_{\rm tot} = \hbar\omega_0\;.  \label{eq:4.5}
\end{align}
Note that if one chooses $u$ to be identical with an angular velocity
\begin{equation}  \label{eq:4.5a}
  u = \omega_0 r\;,
\end{equation}
and with the definition \eqref{eq:2.18} of $\hbar$ so that
\begin{equation}  \label{eq:4.5b}
  \hbar = mur \,,
\end{equation}
one obtains our result \eqref{eq:4.5} immediately from the sub-quantum equipartition rule \eqref{eq:3.9}.

Moreover, if we compare Eq.~(\ref{eq:4.4}) with the Langevin equation (\ref{eq:3.1}), we find the following, additional confirmation of Eq.~(\ref{eq:4.5}). 
First, we recall Boltzmann's relation $\Delta Q=2\omega_0\delta S$ between the heat applied to an oscillating system and a change in the action function $\delta S=\delta\int E_{\rm kin}\d t$, respectively, \cite{Groessing.2008vacuum,Groessing.2009origin} providing
\begin{equation}
	\nabla Q = 2\omega_0 \nabla (\delta S)\;.  \label{eq:4.6}
\end{equation}
$\delta S$ relates to the momentum fluctuation via 
\begin{equation}
  \nabla(\delta S)=\delta\mathbf{p}=: m\mathbf{u} = - \frac{\hbar}{2} \frac{\nabla P}{P}\;,  \label{eq:4.8}
\end{equation}
and therefore, with $P=P_0\e^{-\delta Q/kT_0}$ and \eqref{eq:4.4},
\begin{equation}	\label{eq:4.8a}
	m\mathbf{u} = \frac{\nabla Q}{2\omega_0} \;.
\end{equation}
As the friction force in Eq.~(\ref{eq:3.1}) is equal to the gradient of the heat flux,
\begin{equation} 
   m\zeta\mathbf{u} = \nabla Q \;, \label{eq:4.9}
\end{equation}
comparison of \eqref{eq:4.8a} and \eqref{eq:4.9} provides
\begin{equation}
  \zeta = \gamma = 2\omega_0\;.  \label{eq:4.10}
\end{equation}
Note that with Eqs.~(\ref{eq:4.4a}) and \eqref{eq:4.10} one obtains in any one dimension
the expression for the diffusion constant \eqref{eq:3.6} as
\begin{equation}
	D = \frac{2E_{\rm zp}}{\zeta m} = \frac{\hbar}{2m}\;,  \label{eq:4.12}
\end{equation}
which is exactly the usual expression for $D$ in the context of quantum mechanics.

With Eq.~(\ref{eq:4.8a}) and $\VEC{u}=\omega_0\VEC{r}$ one can also introduce the recently proposed concept of an ``entropic force'' \cite{Verlinde.2010,Padmanabhan.2009}.  That is, with the total energy equaling a total work applied to the system, one can write (with $S_{\rm e}$ denoting the entropy)
\begin{align}
   E_{\rm tot} &= 2 \, \meant{E_{\rm kin}} =: \VEC{F}\cdot\Delta\VEC{x} = T_0\Delta S_{\rm e} = \frac{1}{2\pi}\oint \nabla Q\cdot\d \VEC{r}   \nonumber\\
     &= \Delta Q\,\text{(circle)} = 2\left[\frac{\hbar\omega_0}{4} - \left(-\frac{\hbar\omega_0}{4}\right)\right]
      = \hbar\omega_0  \label{eq:4.13} \;.
\end{align}

Eq.~(\ref{eq:4.13}) provides an ``entropic'' view of a harmonic oscillator in its thermal bath. First, the total energy of a simple harmonic oscillator is given as 
$E_{\rm tot}=mr^2\omega_0^2/2=:\hbar\omega_0/2$. 
Now, the average kinetic energy of a harmonic oscillator is given by half of its total energy, i.e., by 
$\meant{E_{\rm kin}}=mr^2\omega_0^2/4=\hbar\omega_0/4$,
which --- because of the local equilibrium --- is both the average kinetic energy of the bath and that of the ``bouncer'' particle. As the latter during one oscillation varies between $0$ and $\hbar\omega_0/2$, one has the following entropic scenario. When it is minimal, the tendency towards maximal entropy will provide an entropic force equivalent to the absorption of the heat quantity $\Delta Q=\hbar\omega_0/4$. Similarly, when it is maximal, the same tendency will now enforce that the heat $\Delta Q=\hbar\omega_0/4$ is given off again to the ``thermostat'' of the thermal bath. In sum, then, the total energy throughput $E_{\rm tot}$ along a full circle will equal, according to Eq.~(\ref{eq:4.13}), 
$2\meant{E_{\rm kin}}({\rm circle})=2\hbar\omega_0/2=\hbar\omega_0$. In other words, the formula $E=\hbar\omega_0$ does not refer to a classical 
``object'' oscillating with frequency $\omega_0$, but rather to a process of a ``fleeting constancy'': due to entropic requirements, the energy exchange between bouncer and heat bath will constantly consist of absorbing and emitting heat quantities such that in sum the ``total particle energy'' emerges as
$\hbar\omega_0$.

Although the definition \eqref{eq:2.20a} of $\hbar$ indicates an invariant of a particle's dynamics, it still remains to be shown that it is a \textit{universal} invariant, i.e., irrespective of specific particle properties such as $m$ or $\omega_0$, respectively. The universality of $\hbar$ shall be explained in the last chapter, together with the inclusion of spin in our model.

\section{Energy spectrum of the harmonic oscillator from classical physics}

A characteristic and natural feature of nonequilibrium steady-state systems is given by the requirement that the time integral of the so-called
dissipation function $\meant{\Omega_{\rm t}}$ over full periods $\tau$ vanishes identically \cite{Groessing.2008vacuum}. 
With the oscillator's characteristic frequency $\omega_0=2\pi/\tau$, one defines the dissipation function w.r.t.\ the force in Eq.~(\ref{eq:2.1}) over the integral
\begin{equation}
  \frac{1}{\tau}\intlim_0^{\tau}\Omega_{\rm t}\d t 
    := \frac{1}{\tau}\intlim_0^{\tau}\frac{\d F(t)}{kT_0} = 0\;. \label{eq:5.1}
\end{equation}
Here, we assume a generalized driving force $F$ to have a periodic component such that $F(t) \propto \e^{i\omega_0t}$. Then one {\it generally} has that
\begin{equation}
  \intlim_0^{\tau}\d F \propto \e^{i\omega_0(t+\tau)} - \e^{i\omega_0t}\;,  \label{eq:5.2}
\end{equation}
and so the requirement \eqref{eq:5.1} \textit{generally} provides for a whole set of frequencies $\omega_n := n\omega_0 = \frac{2\pi}{\tau_n}$, with $\tau=n\tau_n$, that
\begin{equation}
  \intlim_0^{\tau}\omega_n\d t = 2n\pi\;,\quad\text{for}\; n=1,2,\ldots  \label{eq:5.3}
\end{equation}
(Incidentally, this condition resolves the problem discussed by Wallstrom~\cite{Wallstrom.1994} 
about the single-valuedness of the quantum mechanical wave functions and eliminates
possible contradictions arising from Nelson-type approaches to model quantum
mechanics.)

So, to start with, we are dealing with a situation where a ``particle'' simply oscillates with an angular
frequency $\omega_0$ driven by the external force due to the surrounding (zero-point) fluctuation field, with a
period $\tau=\frac{2\pi}{\omega_0}$. For the type of oscillation we have assumed simple harmonic motion, or, equivalently 
\cite{Feynman.1966vol1}, circular motion, and we have shown in the last paragraph of chapter~\ref{sec:walking} that the total (zero-point) energy is
\begin{equation}
  E_0 = \frac{1}{2}mr^2\omega_0^2 = \frac{\hbar\omega_0}{2}\;.  \label{eq:5.4}
\end{equation}
Then, for slow, adiabatic changes during one period of oscillation,
the action function over a cycle is an invariant,
\begin{equation}
  S_0 = \frac{1}{2\pi}\oint\VEC{p\cdot\d \VEC{r}} = \frac{1}{2\pi}\oint m\omega_0\VEC{r\cdot\d \VEC{r}}\;.  \label{eq:5.5}
\end{equation}
This provides, in accordance with the corresponding standard relation for integrable conservative systems
\cite{Groessing.2008vacuum}, i.e.,
\begin{equation}
  \d S_0 = \frac{\d E_0}{\omega_0}\;,  \label{eq:5.6}
\end{equation}
that
\begin{equation}
  S_0 = \frac{1}{2}mr^2\omega_0\;.  \label{eq:5.7}
\end{equation}

Eq.~\eqref{eq:5.7} can be considered to refer to an ``elementary particle'', i.e., to a simple non-composite mechanical system, which has no excited states. 
More generally, however, the external driving frequency and an arbitrary particle's frequency, respectively, need not be in simple synchrony, since one may have to take into account possible additional energy exchanges of the ``particle'' with its oscillating environment.
Generally, there exists the possibility (within the same boundary condition, i.e., on the circle) of periods $\tau_n=\frac{\tau}{n}=\frac{2\pi}{n\omega_0}=\frac{2\pi}{\omega_n}$, with $n=1,2,\ldots$, of additional adiabatical heat exchanges ``disturbing'' the simple particle oscillation as given by Eq.~\eqref{eq:5.4}.
A concrete example from classical physics is given in \cite{Fort.2010}, where it is shown that a ``path-memory'' w.r.t.\ regular phase-locked wave sources of a ``walker'' in a harmonic oscillator potential can induce the quantization of classical orbits.

Generally, a walker in the experiments of Couder's group is at all times driven by its interactions with a superposition of waves emitted at the points it visited in the past. In other words, the superposition of in-phase waves represents a ``memory'' of a bouncer's path. When disturbed, the trajectory becomes curved. Then, with $n$ disturbances along closed trajectories, only a discrete set of orbits is possible: The path memory is thus proven to be responsible for a self-trapping in quantized orbits, with a corresponding quantization of angular momentum \cite{Fort.2010}.
This scenario is readily transferable to the one discussed in the present paper: Assuming similarly our ``elementary particle'' as a walker on a closed orbit in a harmonic oscillator potential, it may bounce $n$ times during one period $\tau$.

That is, while we have so far considered, via Eqs.~\eqref{eq:5.5} and \eqref{eq:5.6}, a single, slow adiabatic change during an oscillation period, we now also admit the possibility of several (i.e., $n$) additional periodic heat exchanges during the same period, i.e., absorptions and emissions as in \eqref{eq:4.13}. The action integrals over full periods then more generally become 
\begin{equation}
  \oint\d S(\tau_{\rm n}) := -\intlim_0^{\tau}\dot{S}\d t 
	  = \intlim_0^{\tau}E_{\rm tot}\d t
		= \hbar\intlim_0^{\tau}\omega_n\d t\;.  \label{eq:5.8}
\end{equation}
Thus, one can first recall the expressions \eqref{eq:5.4} and \eqref{eq:5.7}, respectively, to obtain for
the case of ``no additional periods'' the basic ``zero-point'' scenario
\begin{equation}
  S_0 = \frac{\hbar}{2}\;, \quad\text{and}\quad E_0 = \frac{1}{2}\hbar\omega_0\;.   \label{eq:5.9}
\end{equation}
Secondly, however, using (\ref{eq:5.3}), one obtains from \eqref{eq:5.8} for $n=1,2,\ldots$ that
\begin{equation}
  \oint\d S(\tau_{\rm n}) = 2n\pi\hbar = nh\;.  \label{eq:5.10}
\end{equation}
This provides a spectrum of $n$ additional possible energy values,
\begin{equation}
  E(n) = \hbar\omega_n = n\hbar\omega_0\;,  \label{eq:5.11}
\end{equation}
such that, together with Eq.~(\ref{eq:5.9}), the total energy spectrum of the off-equilibrium
steady-state harmonic oscillator becomes
\begin{equation} \label{eq:5.12}
  -\frac{\p S}{\p t} = E(n) + E_0
    = \left(n + \frac{1}{2}\right)\hbar\omega_0\;,
    \quad\text{with}\; n=0,1,2,\ldots
\end{equation}
Note that to derive Eq.~(\ref{eq:5.12}) no Schr\"odinger or other quantum mechanical equation was
used. Rather, it was sufficient to invoke Eq.~(\ref{eq:5.2}), without even specifying the exact
expression for $F$.

\section{Introduction of Spin}

Throughout our papers on the sub-quantum thermodynamics of quantum systems \cite{Groessing.2008vacuum,Groessing.2009origin,Groessing.2010emergence,Groessing.2010entropy,Groessing.2010free}, we have made use of the Hamiltonian
\begin{equation}  \label{eq:6.1}
  {\cal H} = \frac{1}{2} m\VEC{v}\cdot\VEC{v} + \frac{1}{2}m\VEC{u}\cdot\VEC{u} + V \;,
\end{equation}
where $V$ is the potential energy and the kinetic energy terms refer to two velocity fields. The latter are denoted as ``convective'' velocity 
\begin{equation}  \label{eq:6.2}
  \VEC{v} := \frac{\VEC{p}}{m} = \frac{\nabla S}{m}
\end{equation}
and ``osmotic'' velocity \cite{Groessing.2008vacuum,Groessing.2009origin,Groessing.2010emergence,Groessing.2010entropy,Groessing.2010free}
\begin{equation}  \label{eq:6.3}
  \VEC{u} := \frac{\delta\VEC{p}}{m} = -D \frac{\nabla P}{P} \;,
\end{equation}
respectively, with a diffusion constant given by Eq.~\eqref{eq:4.12}, $D=\hbar/2m$.
The velocities $\VEC{u}$ and $\VEC{v}$ are by definition irrotational fields, i.e.,
\begin{equation}  \label{eq:6.4}
  \nabla\times\VEC{v} = \frac{1}{m} \nabla\times\nabla S \equiv \VEC{0}
\end{equation}
and
\begin{equation}  \label{eq:6.5}
  \nabla\times\VEC{u} = -\frac{\hbar}{2m} \frac{\nabla\times\nabla P}{P} \equiv \VEC{0} \,.
\end{equation}
Regarding the convective velocity, one usually has the continuity equation
\begin{equation}  \label{eq:6.6}
  \frac{\p}{\p t} P + \nabla\cdot(\VEC{v} P) = 0
\end{equation}
which, with the probability density current
\begin{equation}  \label{eq:6.7}
  \VEC{J} := P \VEC{v} = P \frac{\nabla S}{m} \;,
\end{equation}
is also written as
\begin{equation}  \label{eq:6.8}
  \frac{\p P}{\p t} = - \nabla\cdot\VEC{J} \;.
\end{equation}

However, we are now dealing with a total probability density current
\begin{equation}  \label{eq:6.9}
  \VEC{J} = P (\VEC{v} + \VEC{u}) = \frac{P}{m} (\VEC{p} + \VEC{p}_u) \;,
\end{equation}
where $\VEC{p}=m\VEC{v}$ is the usual ``particle'' momentum and $\VEC{p}_u$ refers to an additional momentum, which is on average orthogonal to it 
\cite{Groessing.2008vacuum,Groessing.2009origin,Groessing.2010emergence}.
How can the conservation of the empirically validated probability current \eqref{eq:6.8} be maintained when the current \eqref{eq:6.7} is extended to include the second term in Eq.~\eqref{eq:6.9}?
The answer was given and discussed by several authors, like in 
\cite{Esposito.1999,Salesi.1996,Recami.1998,Salesi.1998}.
One writes down an ansatz with the introduction of an additional vector $\VEC{s}$,
\begin{equation}  \label{eq:6.10}
  \VEC{u} \to \tilde{\VEC{u}} \times \VEC{s}\;, \quad \text{with} \quad \tilde{\VEC{u}} := \frac{1}{m} \frac{\nabla P}{P} \;,
\end{equation}
such that \eqref{eq:6.9} reads as
\begin{equation}  \label{eq:6.11}
  \VEC{J} =\frac{P}{m} \left( \VEC{p} + \frac{\nabla P}{P} \times \VEC{s} \right) = P (\VEC{v} + \tilde{\VEC{u}} \times \VEC{s}) \;.
\end{equation}
We note that \eqref{eq:6.11}, with the identification of $\VEC{s}$ as a spin vector, is nothing but the non-relativistic limit of the Dirac current, i.e., the Pauli current.
(For a derivation of Eq.\eqref{eq:6.11} in a similar spirit, see \cite{Fritsche.2009}.)
Then, as can easily be shown \cite{Salesi.1996}, Eq.~\eqref{eq:6.8} is fulfilled since, as $\nabla\times\VEC{s}=0$,
\begin{align}  
  \nabla\cdot\VEC{J} &= \nabla\cdot [P(\VEC{v} + \tilde{\VEC{u}} \times \VEC{s})]  %\nonumber\\
    = \nabla\cdot(P\VEC{v}) + \frac{1}{m} \nabla\cdot(\nabla P \times \VEC{s})   \nonumber\\
    &= \nabla\cdot(P\VEC{v}) + \frac{1}{m} \nabla\cdot[\nabla\times (P\VEC{s})] \;.  \label{eq:6.12}
\end{align}
As generally the divergence of a rotation vanishes identically, one re-obtains
\begin{equation}  \label{eq:6.13}
  \nabla\cdot\VEC{J} = \nabla\cdot(P\VEC{v}) = - \frac{\p P}{\p t} \;.
\end{equation}
Note that as $\nabla\times \VEC{s}$ has to vanish, this can be achieved by
\begin{equation}  \label{eq:6.15}
  \hat{\VEC{s}} = \pm\,\VEC{e}_u\times(\VEC{e}_v\times \VEC{e}_u)
\end{equation}
with unit vectors $\VEC{e}_u$ and $\VEC{e}_v$ in the directions of $\VEC{u}$ and $\VEC{v}$, respectively.
Note also that $\hat{\VEC{s}}$ is orthogonal to $\VEC{u}$ and lies in the plane defined by $\VEC{u}$ and $\VEC{v}$.

Finally, with the substitution \eqref{eq:6.10} the Hamiltonian (with $V=0$ for simplicity) reads
\begin{equation}  \label{eq:6.16}
  {\cal H} = \frac{m}{2} (\VEC{v} + \tilde{\VEC{u}}\times\VEC{s})^2 = \frac{m}{2} (v^2 + \tilde{u}^2 s^2) \;,
\end{equation}
which is in agreement with Eqs.~\eqref{eq:6.1} through \eqref{eq:6.3} if and only if $\hat{\VEC{s}}\cdot\hat{\VEC{s}}=1$ and
\begin{equation}  \label{eq:6.17}
  \lvert\VEC{s}\rvert = \frac{\hbar}{2} \;.
\end{equation}

Thus, we see that the two possible vectors $\VEC{s}=\pm \displaystyle\frac{\hbar}{2}\,\hat{\VEC{s}}$ actually do represent the elementary spin of a material particle (fermion) and, comparing with Eq.~\eqref{eq:2.18}, we note that it must be the ``angular momentum generated by the circulating flow of energy in the wave field of the [particle]'' \cite{Yang.2006}. 
In other words, $\VEC{s}$ describes the zitterbewegung of the particle, and one sees that even ``in the Schr\"odinger equation the Planck constant $\hbar$ implicitly denounces the presence of spin'' \cite{Salesi.1996}. 
(Moreover, referring to Eq.~\eqref{eq:2.18}, since also the quantity $n\hbar$ is invariant, with $n=1,2,3,\ldots$, one generally infers the existence of possible spin vector lengths $n\hbar/2$.)
As it is an empirical fact that all fermions are characterized by the same universal spin \eqref{eq:6.17}, we thus conclude that the quantity $\hbar$ defined in \eqref{eq:2.20a} must be universal and thus identical with (the reduced) Planck's constant.

Moreover, we can also provide an additional viewpoint on the quantum potential, in full accordance with C.-D.~Yang's model \cite{Yang.2006} in the complex domain. As in our approach the average quantum potential is given by \cite{Groessing.2008vacuum,Groessing.2009origin}
\begin{equation}  \label{eq:6.18}
  \meanx{U} = \frac{m\meanx{u^2}}{2} = \frac{\hbar\omega_0}{2} \;,
\end{equation}
the quantum potential $U$ can be understood also as an intrinsic torque, which causes the spin angular momentum $\hbar/2$ to precess with an average angular rate $\meanx{\dot{\theta}}=\omega_0$. We thus share the interpretation given by various authors such as 
%\citeauthor{Esposito.1999}, \citeauthor{Salesi.2009}, or \citeauthor{Yang.2006},
Esposito \cite{Esposito.1999}, Recami \cite{Recami.1998}, Salesi \cite{Salesi.1996,Salesi.1998}, or Yang \cite{Yang.2006},
for example, that the existence of spin, the zero-point (or zitterbewegung) oscillations, and the quantum potential are intimately related, even in the non-relativistic framework of the Schr\"odinger equation.

Finally, our model also provides an explanation for the fact that in a measurement the value of only one out of the three spatial spin components can be determined at a specific time. Recall that for our gyrating bouncer to be kept at a constant energy $\hbar\omega_0$, the total energy throughput $E_\text{tot}$ along a full circle, i.e., as discussed via Eq.~\eqref{eq:4.13}, must equal
\begin{equation}  \label{eq:6.19}
  E_\text{tot} = 2\frac{\hbar\omega_0}{2} = 2\lvert \VEC{s}\rvert \omega_0 \;.
\end{equation} 
In other words, the ``entropic view'' presented by Eq.~\eqref{eq:4.13} can be expanded to include the ``spin view'' of the oscillator: during one cycle, it takes up an angular momentum of $\lvert \VEC{s} \rvert=\hbar/2$ and gives off the same amount again, with the net effect of said total throughput, or the ``fleeting constancy'', respectively, of $E_\text{tot}=\hbar\omega_0$. If under such circumstances one fixes one spatial component of $\VEC{s}$, say $s_x$, the requirement of the steady-state system to maintain the energy throughput \eqref{eq:6.19} means that the other components, $s_y$ and $s_z$, must be such that the zero-point energy is still distributed evenly among them.
If it were possible to fix simultaneously more than one of the spin axes, then necessarily the whole mechanism of steady-state maintenance would come to a standstill in all three dimensions. In other words, our off-equilibrium steady-state model ensures that it is not possible to experimentally determine more than one spatial spin component at a time.

%merlin.mbs apsrev4-1.bst 2010-07-25 4.21a (PWD, AO, DPC) hacked
%Control: key (0)
%Control: author (0) dotless jnrlst
%Control: editor formatted (1) identically to author
%Control: production of article title (0) allowed
%Control: page (1) range
%Control: year (0) verbatim
%Control: production of eprint (0) enabled
%

%\bibliography{../../../ains_zotero}  

\end{document}